\documentclass[11pt,a4paper]{article}
\usepackage[english]{babel}
\usepackage{amsmath,amsthm,amssymb,epsfig,latexsym}
\usepackage{color}
\usepackage{ulem}
\usepackage[numbers,square,sort&compress]{natbib}

\newcommand{\so}{\scriptscriptstyle \rm I}
\newcommand{\st}{\scriptscriptstyle \rm I\hspace{-1pt}I}

\newcommand{\la}{u}
\newcommand{\muu}{v}
\newcommand{\lac}{u^{\scriptscriptstyle C}}
\newcommand{\lab}{u^{\scriptscriptstyle B}}
\newcommand{\muc}{v^{\scriptscriptstyle C}}
\newcommand{\mub}{v^{\scriptscriptstyle B}}
\newcommand{\as}{\lambda}
\newcommand{\bla}{\bar u}
\newcommand{\bmu}{\bar v}
\newcommand{\blac}{\bar{u}^{\scriptscriptstyle C}}
\newcommand{\blab}{\bar{u}^{\scriptscriptstyle B}}
\newcommand{\bmuc}{\bar{v}^{\scriptscriptstyle C}}
\newcommand{\bmub}{\bar{v}^{\scriptscriptstyle B}}

\def\Izer{{\sf K}}
\newcommand{\UF}{{\mathfrak{F}}}

\newcommand{\Not}{\mathcal{N}^{(1,2)}}
\newcommand{\Hot}{\mathcal{H}^{(1,2)}}

\newcommand{\Ntt}{\mathcal{N}^{(1,3)}}
\newcommand{\Htt}{\mathcal{H}^{(1,3)}}
\newcommand{\tNot}{\widetilde{\mathcal{N}}^{(1,2)}}
\newcommand{\tHot}{\widetilde{\mathcal{H}}^{(1,2)}}
\newcommand{\tNtt}{\widetilde{\mathcal{M}}}
\newcommand{\tHtt}{\widetilde{\mathcal{H}}^{(1,3)}}

\newcommand{\be}[1]{\begin{equation}\label{#1}}
\newcommand{\ba}[1]{\begin{multline}\label{#1}}
\newcommand{\ee}{\end{equation}}
\newcommand{\ea}{\end{eqnarray}}

\newcommand{\num}{\\\rule{0pt}{20pt}}

\newcommand{\dis}{\displaystyle}

\newcommand{\tr}{\mathop{\rm tr}}

\def\Izer{{\sf K}}

\newtheorem{prop}{Proposition}[section]

 \makeatletter
 \@addtoreset{equation}{section}
 \makeatother
 
\newcommand{\bea}{\begin{eqnarray}}
\newcommand{\eea}{\end{eqnarray}}

\begin{document}

\begin{flushright}
LAPTH-235/14
\end{flushright}

\vspace{20pt}

\begin{center}
\begin{LARGE}
{\bf Zero modes method and form factors in quantum integrable models}
\end{LARGE}

\vspace{40pt}

\begin{large}
{S.~Pakuliak${}^a$, E.~Ragoucy${}^b$, N.~A.~Slavnov${}^c$\footnote{pakuliak@theor.jinr.ru, eric.ragoucy@lapth.cnrs.fr, nslavnov@mi.ras.ru}}
\end{large}

 \vspace{12mm}

\vspace{4mm}

${}^a$ {\it Laboratory of Theoretical Physics, JINR, 141980 Dubna, Moscow reg., Russia,\\
Moscow Institute of Physics and Technology, 141700, Dolgoprudny, Moscow reg., Russia,\\
Institute of Theoretical and Experimental Physics, 117259 Moscow, Russia}

\vspace{4mm}

${}^b$ {\it Laboratoire de Physique Th\'eorique LAPTH, CNRS and Universit\'e de Savoie,\\
BP 110, 74941 Annecy-le-Vieux Cedex, France}

\vspace{4mm}

${}^c$ {\it Steklov Mathematical Institute,
Moscow, Russia}

\end{center}


\vspace{4mm}


\begin{abstract}
We study  integrable models solvable by the nested algebraic Bethe ansatz and possessing
$GL(3)$-invariant $R$-matrix.  Assuming that the monodromy matrix of the model can be expanded
into series with respect to the  inverse spectral parameter, we define zero modes of the monodromy matrix entries
as the first nontrivial coefficients of this series.
Using these
zero modes we establish new relations between form factors of the elements of the monodromy
matrix. We prove that all of them can be obtained from the form factor of a diagonal matrix element
in special limits of Bethe parameters. As a result we obtain determinant representations for form factors of all
the entries of the monodromy matrix.
\end{abstract}

\vspace{1cm}

\vspace{2mm}

\section{Introduction}

The algebraic Bethe ansatz is a powerful method of studying quantum integrable models \cite{FadST79,FadT79,FadLH96,KulR83}.
This method allows one to describe the spectrum of various quantum Hamiltonians in a systematic way. The algebraic Bethe ansatz
also was used for the study of the problem of correlation functions \cite{IzeK84,BogIK93L,KitMST05,Sla07}.
One possible approach to this problem is based on the calculation of form factors of local operators \cite{Sla90,KitMT99,KitKMST09} and
their further summation over the complete set of the Hamiltonian eigenstates \cite{CauM05,CauPS07,KitKMST11,KitKMST12}.

In this paper we continue the study of form factors in $GL(3)$-invariant models, initiated in our previous works \cite{BelPRS12b,BelPRS13a,PakRS14b}.
For a wide class of quantum integrable systems, for which the solution of the quantum inverse scattering problem is known \cite{KitMT99,MaiT00}, the  form factors of local operators can be reduced to the ones of the monodromy matrix entries $T_{ij}(z)$. The calculation of the last ones, in their turn,
reduces to the study of scalar products of Bethe vectors. If one of these vectors is an eigenvector of the quantum Hamiltonian, then for the models
possessing  $GL(2)$ symmetry or its $q$-deformation the corresponding scalar products were calculated in \cite{Sla89}. In this way one can obtain determinant representations for form factors \cite{Sla90,KorKS97,KitMT99}.

For the models with $GL(3)$ symmetry, an analog of the determinant formula obtained in \cite{Sla89} is not known. One should use  a so-called
sum formula for the scalar product of generic Bethe vectors \cite{Res86}. In this representation the scalar product is given as a sum over partitions of Bethe
parameters. In some specific cases this sum can be computed in terms of a single determinant \cite{Whe12,BelPRS12b,BelPRS13a,PakRS14b}. Using this way we
succeeded to find determinant representations for form factors of the operators $T_{ij}(z)$ with $|i-j|\le 1$. However, this straightforward method
of calculation failed in the case of the form factors of the operators $T_{ij}(z)$ with $|i-j|=2$.

In the present paper we develop a new approach to the problem of form factors. It  is applicable  to
 quantum integrable models whose  monodromy matrix $T(z)$
 can be expanded into
a series in the inverse spectral parameter $z^{-1}$ \cite{Dr88,Molevbook}.
We call this approach {\it the  zero
modes method}. In this framework, the form factors of all the operators $T_{ij}(z)$ appear to be related to each other.
We show that if a form factor of a diagonal operator $T_{ii}(z)$ is known, then all other form factors can be obtained from this initial one by sending some of the
Bethe parameters to infinity. The method can be also applied for models with $GL(N)$ symmetry. Here again,
 all other form factors can be obtained from an initial one by sending some of the
Bethe parameters to infinity. However, contrarily to the $GL(3)$ case, this initial form factor remains to be computed.
Yet, some properties can be deduced from the zero modes method.

The article is organized as follows. In section~\ref{S-N} we introduce the model under consideration and describe
the notation used in the paper. We also define the form factors of the monodromy matrix entries  and describe some mappings  between them. In section~\ref{S-ZM} we introduce zero modes of the operators $T_{ij}$ and derive
their action on Bethe vectors. Using these results we find additional relations between the different form factors in
section~\ref{S-RFF}. We show that all the form factors can be obtained from a single initial one by taking special limits of the Bethe parameters.
In section~\ref{S-FFT13} we derive a determinant representation for the  form factor of the monodromy matrix element $T_{13}$. In section~\ref{S-IBR}
we consider a special case when one of the Bethe parameters is infinite.
The generalization to models with $GL(N)$ symmetry is developed in section \ref{sect:glN}.
Appendix~\ref{S-SumF} contains  several summation identities, which are used in section~\ref{S-FFT13} for transformations of determinants.
In appendix~\ref{A-DC} we check relations between different form factors via explicit determinant formulas.

\section{Notation and definitions\label{S-N}}

\subsection{Generalized $GL(3)$-invariant model}

The models considered below are described by the $GL(3)$-invariant
$R$-matrix acting in the tensor product $V_1\otimes V_2$ of two auxiliary spaces
$V_k\sim\mathbb{C}^3$, $k=1,2$:
 \be{R-mat}
 R(x,y)=\mathbf{I}+g(x,y)\mathbf{P},\qquad g(x,y)=\frac{c}{x-y}.
 \ee
In the above definition, $\mathbf{I}$ is the identity matrix in $V_1\otimes V_2$, $\mathbf{P}$ is the permutation matrix
that exchanges $V_1$ and $V_2$, and $c$ is a constant.

The monodromy matrix $T(w)$ satisfies the algebra
\be{RTT}
R_{12}(w_1,w_2)T_1(w_1)T_2(w_2)=T_2(w_2)T_1(w_1)R_{12}(w_1,w_2).
\ee
Equation \eqref{RTT} holds in the tensor product $V_1\otimes V_2\otimes\mathcal{H}$,
where $\mathcal{H}$ is the Hilbert space of the Hamiltonian of the model under consideration.
The  matrices $T_k(w)$ act non-trivially in
$V_k\otimes \mathcal{H}$. Being written in components, equation \eqref{RTT} takes the form
\be{RTT-comp}
\begin{aligned}
{}[T_{ij}(u),T_{kl}(v)] & = g(u,v)\Big(T_{kj}(v)\,T_{il}(u)-T_{kj}(u)\,T_{il}(v)\Big)\\
{}&= g(u,v)\Big(T_{il}(u)\,T_{kj}(v)-T_{il}(v)\,T_{kj}(u)\Big),
\quad i,j,k,l=1,2,3.
\end{aligned}
\ee

The trace in the auxiliary space $V\sim\mathbb{C}^3$ of the monodromy matrix, $\tr T(w)$, is called the transfer matrix. It is a generating
functional of integrals of motion of the model. The eigenvectors of the transfer matrix are
called on-shell Bethe vectors (or simply on-shell vectors). They can be parameterized by sets of complex parameters
satisfying  Bethe equations (see section~\ref{SS-BV}).

Due to the invariance of the
$R$-matrix under transposition with respect to both spaces, the mapping
\be{def-psi}
\psi\,:
T_{ij}(u) \quad\mapsto\quad T_{ji}(u)
\ee
defines an antimorphism of the algebra \eqref{RTT}. One can also prove (see \cite{BelPRS12c}) that
the mapping $\varphi$:
\be{def-phi}
\varphi\,:
T_{ij}(u) \quad\mapsto\quad T_{4-j,4-i}(-u),
\ee
defines an isomorphism of the algebra \eqref{RTT}. The action of the mappings \eqref{def-psi}, \eqref{def-phi}
can be further extended to the action on Bethe vectors and form factors (see sections~\ref{SS-BV},~\ref{SS-FF}).

\subsection{Notation}
We use the same notations and conventions as in the papers \cite{BelPRS13a,PakRS14b}.
Besides the function $g(x,y)$ we also introduce a function $f(x,y)$
\be{univ-not}
 f(x,y)=\frac{x-y+c}{x-y}.
\ee
Two other auxiliary functions  will be also used
\be{desand}
h(x,y)=\frac{f(x,y)}{g(x,y)}=\frac{x-y+c}{c},\qquad  t(x,y)=\frac{g(x,y)}{h(x,y)}=\frac{c^2}{(x-y)(x-y+c)}.
\ee
The following obvious properties of the functions introduced above are useful:
 \be{propert}
 \begin{aligned}
 &g(x,y)\sim \frac cx,\quad &h(x,y)\sim\frac xc,\quad  &f(x,y)\sim 1,\quad  &t(x,y)\sim\frac{c^2}{x^2},\quad &x \to\infty,\\
  &g(x,y)\sim -\frac cy,\quad &h(x,y)\sim-\frac yc,\quad  &f(x,y)\sim 1,\quad  &t(x,y)\sim\frac{c^2}{y^2},\quad &y \to\infty.
 \end{aligned}
 \ee

Before giving a description of the Bethe vectors we formulate a convention on the notations.
We denote sets of variables by bar: $\bar w$, $\bla$, $\bmu$ etc.
Individual elements of the sets are denoted by subscripts: $w_j$, $\la_k$ etc.  Notation $\bla_i$,
means $\bla\setminus u_i$ etc. We say that $\bar x=\bar x'$,
if $\#\bar x=\#\bar x'$ and $x_i=x'_i$ (up to a permutation) for $i=1,\dots,\#\bar x$. We say that $\bar x\ne \bar x'$ otherwise.

In order to avoid too cumbersome formulas we use shorthand notations for products of  operators or
functions depending on one or two variables. Namely, if the functions $g$, $f$, $h$, $t$, as well as the operators $T_{ij}$ depend
on sets of variables, this means that one should take the product over the corresponding set.
For example,
 \be{SH-prod}
 T_{ij}(\bla)=\prod_{\la_k\in\bla} T_{ij}(\la_k);\quad
 g(z, \bar w_i)= \prod_{\substack{w_j\in\bar w\\w_j\ne w_i}} g(z, w_j);\quad
 f(\bla,\bmu)=\prod_{u_j\in\bla}\prod_{v_k\in\bmu} f(u_j,v_k).
 \ee
We emphasize once more that this convention is only valid in the case of functions (or operators), which by definition depend on one or two variables. It does not apply to functions (operators) that depend on  sets of variables.

One of the central object in the study of form factors of $GL(3)$-invariant models is the partition function of the six-vertex model with domain wall boundary conditions (DWPF) \cite{Kor82,Ize87}. We denote it by
$\Izer_n(\bar x|\bar y)$. It depends on two sets of variables $\bar x$ and $\bar y$; the subscript indicates that
$\#\bar x=\#\bar y=n$. The function $\Izer_n$ has the following determinant representation \cite{Ize87}
\begin{equation}\label{K-def}
\Izer_n(\bar x|\bar y)
=\Delta'_n(\bar x)\Delta_n(\bar y)h(\bar x,\bar y)
\det_n t(x_j,y_k),
\end{equation}
where $\Delta'_n(\bar x)$ and $\Delta_n(\bar y)$ are
\be{def-Del}
\Delta'_n(\bar x)
=\prod_{j<k}^n g(x_j,x_k),\qquad {\Delta}_n(\bar y)=\prod_{j>k}^n g(y_j,y_k).
\ee
It is easy to see that $\Izer_n$ is a rational function of $\bar x$ and $\bar y$. If one of its arguments goes to infinity (the other arguments remaining finite),
then the DWPF goes to zero.

\subsection{Bethe vectors\label{SS-BV}}

Now we pass to the description of Bethe vectors.
A generic Bethe vector is denoted by $\mathbb{B}^{a,b}(\bla;\bmu)$.
It is parameterized by two sets of
complex parameters $\bla=\la_1,\dots,\la_a$ and $\bmu=\muu_1,\dots,\muu_b$ with $a,b=0,1,\dots$. They are called Bethe parameters.
Dual Bethe vectors are denoted by $\mathbb{C}^{a,b}(\bla;\bmu)$. They also depend on two sets of
complex parameters $\bla=\la_1,\dots,\la_a$ and $\bmu=\muu_1,\dots,\muu_b$. The state with
$\bla=\bmu=\emptyset$ is called a pseudovacuum vector $|0\rangle$. Similarly the dual state
with $\bla=\bmu=\emptyset$ is called a dual pseudovacuum vector $\langle0|$. These vectors
are annihilated by the operators $T_{ij}(w)$, where $i>j$ for  $|0\rangle$ and $i<j$ for $\langle0|$.
At the same time both vectors are eigenvectors for the diagonal entries of the monodromy matrix
 \be{Tjj}
 T_{ii}(w)|0\rangle=\as_i(w)|0\rangle, \qquad   \langle0|T_{ii}(w)=\as_i(w)\langle0|,\quad i=1,2,3
 \ee
where $\as_i(w)$ are some scalar functions. In the framework of the generalized model, $\as_i(w)$ remain free functional parameters. Actually, it is always possible to normalize
the monodromy matrix $T(w)\to \as_2^{-1}(w)T(w)$ so as to deal only with the ratios
 \be{ratios}
 r_1(w)=\frac{\as_1(w)}{\as_2(w)}, \qquad  r_3(w)=\frac{\as_3(w)}{\as_2(w)}.
 \ee
Below we assume that $\lambda_2(w)=1$.

Different representations for Bethe vectors were found in \cite{TarVar93,KhoPakT07,KhoPak08}.
There exist several explicit formulas for the Bethe vectors in terms of polynomials in $T_{ij}(w)$ (with $i<j$) acting
on the pseudovacuum $|0\rangle$ (see \cite{BelPRS12c}). We give here one of those representations in order to fix
normalization:
\be{BV-expl}
\mathbb{B}^{a,b}(\bla;\bmu) =\sum \frac{\Izer_{k}(\bmu_{\so}|\bla_{\so})}{f(\bmu,\bla)}
f(\bmu_{\st},\bmu_{\so})f(\bla_{\so},\bla_{\st})\,
T_{13}(\bla_{\so})T_{12}(\bla_{\st})T_{23}(\bmu_{\st})|0\rangle.
\ee
Here the sums are taken over partitions of the sets $\bar u\Rightarrow\{\bar u_{\so},\bar u_{\st}\}$ and $\bar v\Rightarrow\{\bar v_{\so}$, $\bar v_{\st}\}$ with $0\leq\#\bar u_{\so}=\#\bar v_{\so}=k\leq\mbox{min}(a,b)$. We recall that the notation $T_{13}(\bla_{\so})$ (and similar ones) means the product of the
operators $T_{13}(u)$ with respect to the subset $\bla_{\so}$.
Finally, $ \Izer_k(\bar v_{\so}|\bar u_{\so})$ is the DWPF \eqref{K-def}.
The normalization used in this formula is the most convenient for deriving explicit formulas of the action of the operators
$T_{ij}(z)$ on $\mathbb{B}^{a,b}(\bla;\bmu)$ \cite{BelPRS12c}.

Dual Bethe vector $\mathbb{C}^{a,b}(\bla;\bmu)$ are related  with $\mathbb{B}^{a,b}(\bla;\bmu)$ by the antimorphysm\footnote{For simplicity we denote mappings
\eqref{def-psi}, \eqref{psiBV}, and \eqref{psiFF} acting on the operators, vectors and form factors by the same letter $\psi$. The same is applied to the mappings \eqref{def-phi}, \eqref{phiBV}, and \eqref{phiFF}.} $\psi$:
\be{psiBV}
\psi \big(\mathbb{B}^{a,b}(\bar u;\bar v)\big) = \mathbb{C}^{a,b}(\bar u;\bar v),
\qquad \psi \big(\mathbb{C}^{a,b}(\bar u;\bar v)\big) = \mathbb{B}^{a,b}(\bar u;\bar v).
\ee
Here we assume that $\psi(|0\rangle) = \langle0|$. Then applying \eqref{def-psi} to \eqref{BV-expl} we obtain
\be{dBV-expl}
\mathbb{C}^{a,b}(\bla;\bmu) =\sum \frac{\Izer_{k}(\bmu_{\so}|\bla_{\so})}{f(\bmu,\bla)}
f(\bmu_{\st},\bmu_{\so})f(\bla_{\so},\bla_{\st})\,
\langle0|T_{32}(\bmu_{\st})T_{21}(\bla_{\st})T_{31}(\bla_{\so}).
\ee

If the parameters $\bla$ and $\bmu$ of a Bethe vector\footnote{%
For simplicity here and below we do not distinguish between vectors and dual vectors.}
satisfy a special system of equations (Bethe equations), then
it becomes an eigenvector of the transfer matrix (on-shell Bethe vector). The system of Bethe equations can be written in the following form:
\be{AEigenS-1}
\begin{aligned}
r_1(u_{i})&=\frac{f(u_{i},\bla_{i})}{f(\bla_{i},u_{i})}f(\bmu,u_{i}),\qquad i=1,\dots,a,\\
r_3(v_{j})&=\frac{f(\bmu_{j},v_{j})}{f(v_{j},\bmu_{j})}f(v_{j},\bla), \qquad j=1,\dots,b.
\end{aligned}
\ee
Recall that $\bla_{i}=\bla\setminus u_i$ and $\bmu_j=\bmu\setminus v_j$.

If $\bla$ and $\bmu$ satisfy the system \eqref{AEigenS-1}, then
\be{Left-act}
\tr T(w)\mathbb{B}^{a,b}(\bla;\bmu) = \tau(w|\bla,\bmu)\,\mathbb{B}^{a,b}(\bla;\bmu),\qquad
\mathbb{C}^{a,b}(\bla;\bmu)\tr T(w) = \tau(w|\bla,\bmu)\,\mathbb{C}^{a,b}(\bla;\bmu),
\ee
where
\be{tau-def}
\tau(w)\equiv\tau(w|\bla,\bmu)=r_1(w)f(\bla,w)+f(w,\bla)f(\bmu,w)+r_3(w)f(w,\bmu).
\ee

 {\sl Remark.\label{GenMod1}} In concrete quantum models the functions $r_1(w)$ and $r_3(w)$ are fixed. Then the system of Bethe equations 
\eqref{AEigenS-1} determines the admissible values of the parameters $\bla$ and $\bmu$. Eventually these values characterize the spectrum
of the Hamiltonian of the quantum model under consideration.  However, in the generalized model, where  $r_1(w)$ and $r_3(w)$ are free functional
parameters, the situation is opposite. The system \eqref{AEigenS-1} only fixes the values of the functions $r_1(w)$ and $r_3(w)$ in several
points, while the parameters $\bla$ and $\bmu$ remain arbitrary complex numbers \cite{Kor82}.

\subsection{Form factors of the monodromy matrix entries\label{SS-FF}}

Form factors of the monodromy matrix entries are defined as
 \be{SP-deFF-gen}
 \mathcal{F}_{a,b}^{(i,j)}(z)\equiv\mathcal{F}_{a,b}^{(i,j)}(z|\blac,\bmuc;\blab,\bmub)=
 \mathbb{C}^{a',b'}(\blac;\bmuc)T_{ij}(z)\mathbb{B}^{a,b}(\blab;\bmub),
 \ee
where both $\mathbb{C}^{a',b'}(\blac;\bmuc)$ and $\mathbb{B}^{a,b}(\blab;\bmub)$ are on-shell
Bethe vectors, and
\be{apabpb}
\begin{array}{l}
a'=a+\delta_{i1}-\delta_{j1},\\
b'=b+\delta_{j3}-\delta_{i3}.
\end{array}
\ee
The parameter $z$ is an arbitrary complex  number. We call it  the external parameter.

Obviously, there exist nine form factors of $T_{ij}(z)$ in the models with $GL(3)$-invariant
$R$-matrix. However, not all of them are independent. In particular, due to the mapping \eqref{def-psi}
one can easily show that
\be{psiFF}
\psi \big(\mathcal{F}_{a,b}^{(i,j)}(z|\blac,\bmuc;\blab,\bmub)\big) =
\mathcal{F}_{a',b'}^{(j,i)}(z|\blab,\bmub;\blac,\bmuc),
\ee
and hence, the
form factor $\mathcal{F}_{a,b}^{(i,j)}(z)$ can be obtained from  $\mathcal{F}_{a,b}^{(j,i)}(z)$
via the replacements of the Bethe parameters $\{\blac,\bmuc\}\leftrightarrow\{\blab,\bmub\}$ and the cardinalities of the sets $\{a,b\}\leftrightarrow\{a',b'\}$.

One more relationship between different form factors appears due to
the isomorphism \eqref{def-phi}, that implies the following transform of Bethe vectors:
\be{phiBV}
\varphi \big(\mathbb{B}^{a,b}(\bar u;\bar v)\big) = \mathbb{B}^{b,a}(-\bar v;-\bar u),
\qquad \varphi \big(\mathbb{C}^{a,b}(\bar u;\bar v)\big) = \mathbb{C}^{b,a}(-\bar v;-\bar u).
\ee
Since the mapping $\varphi$ connects the operators $T_ {11}$ and $T_ {33}$, it also leads to the replacement of functions
$r_1 \leftrightarrow r_3$.
Therefore, if $\mathbb{B}^{a,b}(\bar u;\bar v)$ and $\mathbb{C}^{a,b}(\bar u;\bar v)$ are constructed in the representation
$\mathcal{V}\big(r_1(u),r_3(u)\big)$, when their images are in the
representation $\mathcal{V}\big(r_3(-u),r_1(-u)\big)$.  Hence, we obtain one more relation for form factors
\be{phiFF}
\varphi \big(\mathcal{F}_{a,b}^{(i,j)}(z|\blac,\bmuc;\blab,\bmub)\big) =
\mathcal{F}_{b,a}^{(4-j,4-i)}(-z|-\bmuc,-\blac;-\bmub,-\blab)\Bigr|_{r_1\leftrightarrow r_3}.
\ee

Thus, it is enough to calculate only four form factors, say, $\mathcal{F}^{(1,1)}(z)$, $\mathcal{F}^{(1,2)}(z)$, $\mathcal{F}^{(1,3)}(z)$
and $\mathcal{F}^{(2,2)}(z)$. All others can be obtained from these four by the mappings $\psi$ and $\varphi$.

\section{Zero modes\label{S-ZM}}

Assume that monodromy matrix $T(u)$ can be expanded  into a series over $u^{-1}$ of the form:
\be{zero-modes}
T_{ij}(u)=\delta_{ij}+ \sum_{n=0}^\infty T_{ij}[n]\,\left(\tfrac cu\right)^{n+1}.
\ee
{This may happen if monodromy matrix of the model is obtained as specialization to some highest weight representation
of the Yangian   $Y(\mathfrak{gl}_3)$ with highest weight vector $|0\rangle$ \cite{Dr88,Molevbook}.}

Note that the expansion \eqref{zero-modes} yields similar expansions for the functions $r_1(u)$ and $r_3(u)$
\be{zero-modes-r}
r_k(u)=1+ \sum_{n=0}^\infty r_k[n]\,\left(\tfrac cu\right)^{n+1}.
\ee
Assumption \eqref{zero-modes} implies that the Bethe vectors remain on-shell if one of their parameters
tends to infinity.
This is because the structure of the Bethe equations \eqref{AEigenS-1} is preserved
 when $r_k(u)\to 1$ at $u\to\infty$.

The operators $T_{ij}[0]$ are called the zero modes. They generate the $GL(3)$ algebra that is a symmetry of the model and
play a very important role in our further considerations.
Sending in \eqref{RTT-comp} one of the arguments to infinity we obtain
\be{RTT-zero}
[T_{ij}[0], T_{kl}(u)]  =  \delta_{il}T_{kj}(u)-\delta_{kj}T_{il}(u).
\ee

\subsection{Action of the  zero modes onto Bethe vectors}

The explicit formulas for the action the operators $T_{ij}(z)$ onto Bethe vectors were derived in \cite{BelPRS12c}. Taking
the limit $z\to\infty$ in those expressions we obtain the action of zero modes $T_{ij}[0]$. The action of $T_{ij}[0]$ with
$i<j$ is given by
\begin{align}\label{act-13}
T_{13}[0]\mathbb{B}^{a,b}(\bla;\bmu)&=
\lim_{w\to\infty} \tfrac wc\; \mathbb{B}^{a+1,b+1}(\{\bla,w\};\{\bmu,w\}),\\
\label{act-12}
T_{12}[0]\mathbb{B}^{a,b}(\bla;\bmu)&=
\lim_{w\to\infty} \tfrac wc\; \mathbb{B}^{a+1,b}(\{\bla,w\};\bmu),\\
\label{act-23}
T_{23}[0]\mathbb{B}^{a,b}(\bla;\bmu)&=
\lim_{w\to\infty} \tfrac wc\; \mathbb{B}^{a,b+1}(\bla;\{\bmu,w\}).
\end{align}
Observe that due to the normalization used in the expression  \eqref{BV-expl}, the Bethe vector goes to zero if one of its arguments goes to infinity.
Multiplication by $w$ like in \eqref{act-13}--\eqref{act-23} makes the result finite.
The parameters $\bla$ and $\bmu$ in \eqref{act-13}--\eqref{act-23} are a priori generic
complex numbers, but they may satisfy  the Bethe equations in specific cases. Then in the r.h.s. of \eqref{act-12} and \eqref{act-23} we
obtain on-shell Bethe vectors, because the infinite root $w$ together with the sets $\bla$ and $\bmu$ satisfy Bethe equations due to the
condition \eqref{zero-modes-r}.

The action of the diagonal zero modes takes the following form:
\begin{align}
T_{11}[0]\mathbb{B}^{a,b}(\bla;\bmu)&=(r_1[0]-a)\mathbb{B}^{a,b}(\bla;\bmu),\label{act-11} \\
T_{22}[0]\mathbb{B}^{a,b}(\bla;\bmu)&=(a-b)\mathbb{B}^{a,b}(\bla;\bmu),\label{act-22} \\
T_{33}[0]\mathbb{B}^{a,b}(\bla;\bmu)&=(r_3[0]+b)\;\mathbb{B}^{a,b}(\bla;\bmu). \label{act-33}
\end{align}
Thus, a generic Bethe vector $\mathbb{B}^{a,b}(\bla;\bmu)$ is an eigenvector of the diagonal zero modes $T_{ii}[0]$.

Finally, the action of the zero modes $T_{ij}[0]$ with
$i>j$ is a bit more complex. We first present this action in the case when the parameters $\bla$ and $\bmu$ are finite. Then
\begin{align}\label{act-21}
T_{21}[0]\mathbb{B}^{a,b}(\bla;\bmu)=
\sum_{i=1}^a\Bigl\{\frac{r_1(u_i)f(\bla_i,u_i)}{f(\bmu,u_i)}-f(u_i,\bla_i)\Bigr\}\mathbb{B}^{a-1,b}(\bla_i;\bmu),\\
T_{32}[0]\mathbb{B}^{a,b}(\bla;\bmu)=
-\sum_{i=1}^b\Bigl\{\frac{r_3(v_i)f(v_i,\bmu_i)}{f(v_i,\bla)}-f(\bmu_i,v_i)\Bigr\}\mathbb{B}^{a,b-1}(\bla;\bmu_i). \label{act-32}
\end{align}
We do not give here the action of $T_{31}[0]$ because it is more cumbersome and we do not use it below. Observe that if
$\mathbb{B}^{a,b}(\bla;\bmu)$ is an on-shell vector, then the r.h.s. of \eqref{act-21}, \eqref{act-32} vanish due to the Bethe equations
\eqref{AEigenS-1}. Thus, the on-shell vectors depending on finite Bethe roots are  {\it singular weight} vectors of the zero modes  $T_{ij}[0]$
with\footnote{Due to commutation relation \eqref{RTT-zero}, singularity of the on-shell Bethe vectors with respect to the zero mode
$T_{31}[0]$ follows from \eqref{act-21}, \eqref{act-32} and   the commutation relation $T_{31}[0]=[T_{21}[0],T_{32}[0]]$.}
$i>j$ (see also \cite{MuhTV06} for $GL(N)$ case).
The case when one of the Bethe roots is infinite will be considered in section~\ref{S-IBR}.

The action of the zero modes on the dual vectors $\mathbb{C}^{a,b}(\bla;\bmu)$ can be obtained by the antimorphysm $\psi$ \eqref{psiBV}.
In particular,
\begin{align}
\label{act-21d}
\mathbb{C}^{a,b}(\bla;\bmu)T_{21}[0]&=
\lim_{w\to\infty} \tfrac wc\; \mathbb{C}^{a+1,b}(\{\bla,w\};\bmu),\\
\label{act-32d}
\mathbb{C}^{a,b}(\bla;\bmu)T_{32}[0]&=
\lim_{w\to\infty} \tfrac wc\; \mathbb{C}^{a,b+1}(\bla;\{\bmu,w\}),
\end{align}
and
\begin{equation}
\label{act-12-23}
\mathbb{C}^{a,b}(\bla;\bmu)T_{12}[0]=0,\qquad
\mathbb{C}^{a,b}(\bla;\bmu)T_{23}[0]=0,
\end{equation}
if $\mathbb{C}^{a,b}(\bla;\bmu)$ is an on-shell Bethe vector depending on finite parameters.

\section{Relations between different form factors\label{S-RFF}}

Setting in \eqref{RTT-zero} $i=l=2$, $j=3$, and $k=1$ we obtain
\be{RTT-exp}
[T_{23}[0],T_{12}(z)]=T_{13}(z).
\ee

Let $\mathbb{C}^{a+1,b+1}(\blac;\bmuc)$ and $\mathbb{B}^{a,b}(\blab;\bmub)$ be two on-shell vectors with
all  Bethe parameters finite.  Then \eqref{RTT-exp} yields
\begin{multline}
\mathbb{C}^{a+1,b+1}(\blac;\bmuc) T_{13}(z)\mathbb{B}^{a,b}(\blab;\bmub)=
\mathbb{C}^{a+1,b+1}(\blac;\bmuc) T_{23}[0]T_{12}(z)\mathbb{B}^{a,b}(\blab;\bmub) \\
-\mathbb{C}^{a+1,b+1}(\blac;\bmuc) T_{12}(z)T_{23}[0]\mathbb{B}^{a,b}(\blab;\bmub) .\label{T13-z1}
\end{multline}
The first term in the r.h.s. vanishes as $T_{23}[0]$ acts on the dual on-shell Bethe vector. The action of
$T_{23}[0]$ on the on-shell vector $\mathbb{B}^{a,b}(\blab;\bmub)$  is given by \eqref{act-23}, hence,
\begin{equation}
\mathbb{C}^{a+1,b+1}(\blac;\bmuc) T_{13}(z)\mathbb{B}^{a,b}(\blab;\bmub)=
-\mathbb{C}^{a+1,b+1}(\blac;\bmuc) T_{12}(z)\lim_{w\to\infty}\tfrac{w}c\;\mathbb{B}^{a,b+1}(\blab;\{\bmub,w\}) .\label{T13-z2}
\end{equation}
Since the original vector $\mathbb{B}^{a,b}(\blab;\bmub)$ was on-shell, the new vector  $\mathbb{B}^{a,b+1}(\blab;\{\bmub,w\})$ with $w\to\infty$ also is on-shell. Thus, in the r.h.s.
of \eqref{T13-z2} we have the form factor of $T_{12}(z)$, and we arrive at
\be{FF-FF}
\mathcal{F}^{(1,3)}_{a,b}(z|\blac,\bmuc;\blab,\bmub)=-\lim_{w\to\infty}\tfrac{w}c\;
\mathcal{F}^{(1,2)}_{a,b+1}(z|\blac,\bmuc;\blab,\{\bmub,w\}).
\ee

Similarly one can obtain relations between other form factors. In particular, setting in \eqref{RTT-zero} $i=1$, $j=2$, and $k=l=\epsilon$
($\epsilon=1,2$) we obtain
\be{FF-FF-n}
\mathcal{F}^{(1,2)}_{a,b}(z|\blac,\bmuc;\blab,\bmub)=(-1)^\epsilon\lim_{w\to\infty}\tfrac{w}c\;
\mathcal{F}^{(\epsilon,\epsilon)}_{a+1,b}(z|\blac,\bmuc;\{\blab,w\},\bmub),\qquad \epsilon=1,2.
\ee
Finally, setting in \eqref{RTT-zero} $i=l=2$, $j=k=1$, we find
\be{F11F22-F12}
\mathcal{F}^{(1,1)}_{a,b}(z|\blac,\bmuc;\blab,\bmub)-
\mathcal{F}^{(2,2)}_{a,b}(z|\blac,\bmuc;\blab,\bmub)
=\lim_{w\to\infty}\tfrac{w}c\;
\mathcal{F}^{(1,2)}_{a,b}(z|\{\blac,w\},\bmuc;\blab,\bmub).
\ee

Thus, we arrive at the following
\begin{prop}\label{all-one}
All form factors in the $GL(3)$-invariant generalized model can be obtained from only one form factor by sending one of Bethe parameters to infinity.
\end{prop}

Indeed, we can begin, for instance, with the form factor $\mathcal{F}^{(2,2)}_{a,b}(z)$. Using
\eqref{FF-FF-n} we obtain $\mathcal{F}^{(1,2)}_{a,b}(z)$. Then applying \eqref{F11F22-F12} and
\eqref{FF-FF} we respectively find the form factors $\mathcal{F}^{(1,1)}_{a,b}(z)$ and
$\mathcal{F}^{(1,3)}_{a,b}(z)$. All other form factors can be obtained via the mappings $\psi$ \eqref{psiFF} and $\phi$ \eqref{phiFF},
but it is clear that one can also find these form factors starting from $\mathcal{F}^{(2,2)}_{a,b}(z)$ and taking special limits of
the Bethe parameters. In its turn, the calculation of the initial form factor $\mathcal{F}^{(2,2)}_{a,b}(z)$ reduces to the
calculation of the scalar product of twisted on-shell and usual on-shell Bethe vectors \cite{BelPRS12b,BelPRS13a}.

{\sl Remark.} The commutation relations \eqref{RTT-zero} also hold in the $GL(N)$-invariant generalized model with $N>3$. Therefore
one can derive the relations of the type \eqref{FF-FF}--\eqref{F11F22-F12} for this model and prove that all form factors
of the monodromy matrix entries $T_{ij}(z)$ follow from an initial form factor of  a diagonal element. We briefly describe the $GL(N)$ case
in section~\ref{sect:glN}.

Explicit determinant formulas for form factors $\mathcal{F}^{(2,2)}_{a,b}(z)$, $\mathcal{F}^{(1,1)}_{a,b}(z)$, and
$\mathcal{F}^{(1,2)}_{a,b}(z)$ in $GL(3)$-invariant generalized model were obtained in \cite{BelPRS12b,BelPRS13a,PakRS14b}.
Those formulas were derived by a straightforward method based on a representation for the scalar product of
Bethe vectors \cite{Res86}. Using  explicit determinant representations for the form factors listed above one can convince
 himself that equations \eqref{FF-FF}--\eqref{F11F22-F12} indeed are valid.

It should be noted that the possibility of considering the limit of an infinite Bethe parameter is based on the use of the generalized model. On the one hand, in this model, the Bethe parameters  are arbitrary complex numbers. Hence, one of them can be sent to infinity. On the other
hand, the existence  of an infinite root in the Bethe equations agrees  with the expansion \eqref{zero-modes-r}. At the same time, the condition
\eqref{zero-modes-r} is not a restriction of the free functional parameters  $r_1$  and  $r_3$, since it is not used  in calculating the form factor limits. This explains the fact that the determinant representation for
the  form factors $\mathcal{F}^{(i,j)}_{a,b}(z)$ with $|i-j|\le 1$ satisfy conditions \eqref{FF-FF}--\eqref{F11F22-F12}, despite  these representations were obtained without any additional assumptions on the behavior of the  functions $r_1$  and  $r_3$ at infinity.

As we have mentioned already, the straightforward method of calculation failed in the case of the form factor
$\mathcal{F}^{(i,j)}_{a,b}(z)$ with $|i-j|=2$ , and thus, determinant representations for these form factors were not known up to now. Equation \eqref{FF-FF} allows one to solve this problem in a simple way for $\mathcal{F}^{(1,3)}_{a,b}(z)$. Knowing a representation for the form factor $\mathcal{F}^{(1,3)}_{a,b}(z)$ we can easily obtain
 one for $\mathcal{F}^{(3,1)}_{a,b}(z)$ via the mapping \eqref{psiFF}.  We will detail this question in section~\ref{S-FFT13}.

Note that Proposition~\ref{all-one} allows us to find explicitly the dependence on the external parameter $z$ for all form factors.
\begin{prop}
Given sets $\blac$, $\blab$, $\bmuc$, and $\bmub$ assume that $\blac\ne\blab$ or $\bmuc\ne\bmub$. Then for all form factors
$\mathcal{F}^{(\epsilon,\epsilon')}_{a,b}(z|\blac,\bmuc;\blab,\bmub)$, $\epsilon,\epsilon'=1,2,3$, the dependence on the external parameter $z$ is given by
\be{dep-z}
\mathcal{F}^{(\epsilon,\epsilon')}_{a,b}(z|\blac,\bmuc;\blab,\bmub)=\bigl(\tau(z|\blac,\bmuc)-\tau(z|\blab,\bmub)\bigr)\cdot
\UF^{(\epsilon,\epsilon')}_{a,b}(\blac,\bmuc;\blab,\bmub),
\ee
where $\tau(z|\bla,\bmu)$ is the transfer matrix eigenvalue \eqref{tau-def}, and
$\UF^{(\epsilon,\epsilon')}_{a,b}(\blac,\bmuc;\blab,\bmub)$ does not depend on $z$. We call
$\UF^{(\epsilon,\epsilon')}_{a,b}(\blac,\bmuc;\blab,\bmub)$ a  universal form factor,  because
it  is determined by the $R$-matrix only, and does  not depend on the functions $r_k$ which specify a quantum model.
\end{prop}
 {\sl Remark.\label{GenMod2}} Strictly speaking the universal form factor does not depend on a concrete model, if 
$\blac\cap\blab=\emptyset$ and $\bmuc\cap\bmub=\emptyset$. Otherwise it depends on the derivatives of the functions
$r_k$. We consider this case in section~\ref{SS-FFIBR}.

{\sl Proof.}  It was proved in \cite{BelPRS13a} that equation \eqref{dep-z} holds at least for the form factors of the diagonal entries $T_{ii}(z)$.
In particular,
\be{dep-z22}
\mathcal{F}^{(2,2)}_{a,b}(z|\blac,\bmuc;\blab,\bmub)=\bigl(\tau(z|\blac,\bmuc)-\tau(z|\blab,\bmub)\bigr)\cdot
\UF^{(2,2)}_{a,b}(\blac,\bmuc;\blab,\bmub),
\ee
where $\UF^{(2,2)}_{a,b}(\blac,\bmuc;\blab,\bmub)$ does not depend on $z$.  We know that all other
form factors are special limits of $\mathcal{F}^{(2,2)}_{a,b}(z|\blac,\bmuc;\blab,\bmub)$, where one of the Bethe parameters
goes to infinity. Looking at the explicit expression \eqref{tau-def} for the eigenvalue $\tau(z|\bla,\bmu)$ we see that
\be{lim-tau}
\lim_{u_a\to\infty}\tau(z|\bla,\bmu)=\tau(z|\bla_a,\bmu),\qquad
\lim_{v_b\to\infty}\tau(z|\bla,\bmu)=\tau(z|\bla,\bmu_b).
\ee
Thus, if one of the Bethe parameters goes to infinity, then the transfer matrix eigenvalue $\tau(z|\bla,\bmu)$ turns into the
eigenvalue depending on the remaining Bethe parameters. Hence, the structure \eqref{dep-z22} is preserved in all the limiting cases.

Note that equation \eqref{dep-z} also can be proved by means of explicit determinant representations for form factors.

\section{Form factor of $T_{13}$\label{S-FFT13}}

In this section we obtain a determinant representation for the form factor of the operator $T_{13}(z)$. Recall that
in the form factor $\mathcal{F}^{(1,3)}_{a,b}(z|\blac,\bmuc;\blab,\bmub)$ the cardinalities of the Bethe parameters are
\be{card}
\#\blab=a,\qquad \#\blac=a+1,\qquad \#\bmub=b,\qquad  \#\bmuc=b+1.
\ee
To describe the determinant formula we introduce a set $\bar x'=\{x'_1,\dots,x'_{a+b+1}\}$ as a union of the sets $\blab$
and $\bmuc$:  $\bar x'=\{\blab,\bmuc\}$.
Let
\be{hHab}
\mathcal{H}^{(1,3)}_{a,b}= \frac{h(\bar x',\blab)h(\bmuc,\bar x')}{h(\bmuc,\blab)}\;\Delta'_{a+1}(\blac)
\Delta'_{b}(\bmub)\Delta_{a+b+1}(\bar x'),
\ee
where $h$ is defined in \eqref{desand}  and $\Delta'$, $\Delta$ are given by \eqref{def-Del}.
The subscripts $a+1$ and $b$ of this function are equal to the cardinalities of the sets $\blac$ and $\bmub$ respectively.

\begin{prop}
The form factor $\mathcal{F}^{(1,3)}_{a,b}(z|\blac,\bmuc;\blab,\bmub)$ admits the following determinant representation:
\be{FF-ans-13-mod}
\mathcal{F}^{(1,3)}_{a,b}(z|\blac,\bmuc;\blab,\bmub)=\bigl(\tau(z|\blac,\bmuc)-\tau(z|\blab,\bmub)\bigr)
\mathcal{H}^{(1,3)}_{a,b}\;
\det_{a+b+1}\bigl(\Ntt_{jk}\bigr),
\ee
where the eigenvalue of the transfer matrix $\tau(z|\bla,\bmu)$ is given by \eqref{tau-def}. The entries
of the matrix $\Ntt$ have the following form:
\be{Lu-13}
\mathcal{N}^{(1,3)}_{jk}=t(\lac_j,x'_k)\frac{(-1)^{a}r_1(x'_k)h(\blac,x'_k)}{f(\bmuc,x'_k)h(x'_k,\blab)}
+t(x'_k,\lac_j)\frac{h(x'_k,\blac)}{h(x'_k,\blab)}, \qquad \begin{array}{l}j=1,\dots, a+1,\\
k=1,\dots,a+b+1,\end{array}
\ee
and
\be{Lv-13}
\Ntt_{a+1+j,k}=t(x'_k,\mub_j)\frac{(-1)^{b-1}r_3(x'_k)h(x'_k,\bmub)}{f(x'_k,\blab)h(\bmuc,x'_k)}
+t(\mub_j,x'_k)\frac{h(\bmub,x'_k)}{h(\bmuc,x'_k)}, \qquad \begin{array}{l}j=1,\dots, b,\\
k=1,\dots,a+b+1.\end{array}
\ee
\end{prop}

{\sl Proof.} Due to equation \eqref{FF-FF} the form factor $\mathcal{F}^{(1,3)}_{a,b+1}(z)$ is equal to the
limit of the form factor $\mathcal{F}^{(1,2)}_{a,b+1}(z)$  where one of the Bethe parameters goes to infinity. Hence, in order to prove
representation \eqref{FF-ans-13-mod} it is enough to take this limit in the determinant formula for $\mathcal{F}^{(1,2)}_{a,b+1}(z)$,  obtained in \cite{PakRS14b}, that we recall below.

We introduce a set of variables $\bar x=\{x_1,\dots,x_{a+b+2}\}$ as the union of  the sets
\be{Set-x}
\bar x=\{\blab,\bmuc,z\}=\{\lab_1,\dots,\lab_a,\muc_1,\dots,\muc_{b+1},z\}\,.
\ee
Then
\be{FF-ans-m}
\mathcal{F}^{(1,2)}_{a,b+1}(z|\blac,\bmuc;\blab,\{\bmub,w\})=\Hot_{a,b+1}\;
\det_{a+b+2}\bigl(\mathcal{N}^{(1,2)}_{jk}\bigr).
\ee
Here the coefficient $\Hot_{a,b+1}$  has the form
\be{Hab}
\Hot_{a,b+1}=\frac{h(\bar x,\blab)h(\bmuc,\bar x)}{h(\bmuc,\blab)}\;\Delta'_{a+1}(\blac)
\Delta'_{b+1}(\{\bmub,w\})\Delta_{a+b+2}(\bar x).
\ee
The subscripts $a+1$ and $b+1$ denote the cardinalities of the sets $\blac$ and $\{\bmub,w\}$ respectively.
The matrix $\mathcal{N}^{(1,2)}_{jk}$ consists of three blocks.  For $k=1,\dots,a+b+2$ one has
\be{Lu-m}
\mathcal{N}^{(1,2)}_{jk}=t(\lac_j,x_k)\frac{(-1)^{a}r_1(x_k)h(\blac,x_k)}{f(\bmuc,x_k)h(x_k,\blab)}
+t(x_k,\lac_j)\frac{h(x_k,\blac)}{h(x_k,\blab)}, \qquad j=1,\dots, a+1,
\ee
\begin{multline}\label{Lv-m}
\mathcal{N}^{(1,2)}_{j+a+1,k}=t(x_k,\mub_j)\frac{(-1)^{b}r_3(x_k)h(x_k,\bmub)h(x_k,w)}{f(x_k,\blab)h(\bmuc,x_k)}\\
+t(\mub_j,x_k)\frac{h(\bmub,x_k)h(w,x_k)}{h(\bmuc,x_k)},\qquad j=1,\dots, b,
\end{multline}
and
\begin{equation}\label{Lw-m}
\mathcal{N}^{(1,2)}_{a+b+2,k}=g(x_k,w)\frac{(-1)^{b}r_3(x_k)h(x_k,\bmub)}{f(x_k,\blab)h(\bmuc,x_k)}
+g(w,x_k)\frac{h(\bmub,x_k)}{h(\bmuc,x_k)}.
\end{equation}

It is convenient to introduce
\be{tHab}
\tHot_{a,b+1}=\left(\tfrac{w}c\right)^b \Hot_{a,b+1},
\ee
and  for all $k=1,\dots,a+b+2$,
\be{tLu-m}
\begin{aligned}
&\tNot_{jk}=\mathcal{N}^{(1,2)}_{jk} \qquad&j=1,\dots, a+1,\\
&\tNot_{a+1+j,k}=\tfrac{c}{w}\;\mathcal{N}^{(1,2)}_{a+1+j,k}\qquad&j=1,\dots,  b,\\
&\tNot_{a+b+2,k}=\tfrac{w}{c}\;\mathcal{N}^{(1,2)}_{a+b+2,k}.\qquad&{}
\end{aligned}
\ee
Then due to \eqref{FF-FF} we have
\be{F13-tNot}
\mathcal{F}^{(1,3)}_{a,b}(z|\blac,\bmuc;\blab,\bmub)=-\lim_{w\to\infty}\tHot_{a,b+1}\;
\det_{a+b+2}\bigl(\tNot_{jk}\bigr).
\ee

Consider the limit $w\to\infty$ of the prefactor $\tHot_{a,b+1}$. Here  only the function $\Delta'_{b+1}(\{\bmub,w\})$
depends on $w$. Using \eqref{propert}, \eqref{def-Del} we obtain
\be{Hab-lim}
\lim_{w\to\infty}\tHot_{a,b+1}=(-1)^b \frac{h(\bar x,\blab)h(\bmuc,\bar x)}{h(\bmuc,\blab)}\;\Delta'_{a+1}(\blac)
\Delta'_{b}(\bmub)\Delta_{a+b+2}(\bar x).
\ee
Let us extract explicitly in \eqref{Hab-lim} the dependence on
the external parameter $z$.  Recall that $\bar x=\{\bar x',z\}$, where $\bar x'=\{\blab,\bmuc\}$. Then obviously
\be{Extract-z}
\begin{aligned}
&h(\bar x,\blab)h(\bmuc,\bar x)=h(z,\blab)h(\bmuc,z)\cdot h(\bar x',\blab)h(\bmuc,\bar x'),\\
&\Delta_{a+b+2}(\bar x)=g(z,\blab)g(z,\bmuc)\cdot \Delta_{a+b+1}(\bar x'),
\end{aligned}
\ee
and using \eqref{desand},  \eqref{hHab} we find
\be{Hab-lim1}
\lim_{w\to\infty}\tHot_{a,b+1}=f(z,\blab)f(\bmuc,z)\;\Htt_{a,b}.
\ee

Let us pass now to the limit of the matrix $\tNot$.
The entries $\tNot_{jk}$ with $j\le a+1$ do not depend on $w$, therefore they do not change in
the limit $w\to\infty$. Comparing these matrix elements with $\Ntt_{jk}$ \eqref{Lv-13} we see  that $\tNot_{jk}=\Ntt_{jk}$ for $j\le a+1$
and $k=1,\dots,a+b+1$. In the last column we have
\be{last-col-u}
\lim_{w\to\infty}\tNot_{j,a+b+2}=\Ntt_{jk}\Bigr|_{x'_k=z},\qquad j=1,\dots,a+1.
\ee

Consider now the limit of the entries $\tNot_{a+1+j,k}$ for $j<b+1$. Using \eqref{propert} one can easily see that
\be{lim-M}
\begin{aligned}
&\lim_{w\to\infty}\tNot_{a+1+j,k}=\Ntt_{a+1+j,k},\qquad j=1,\dots, b,\qquad k=1,\dots,a+b+1,\\
&\lim_{w\to\infty}\tNot_{a+1+j,a+b+2}=\Ntt_{a+1+j,k}\Bigr|_{x'_k=z},\qquad j=1,\dots, b,
\end{aligned}
\ee

where $\Ntt_{a+1+j,k}$ is given by \eqref{Lv-13}.
Finally, in the last row of the matrix $\tNot_{j,k}$ we obtain
\be{lim-Notlast}
\lim_{w\to\infty}\tNot_{a+b+2,k}=\Phi_{a,b}(x_k),\qquad k=1,\dots,a+b+2,
\ee
where
\be{phi-k}
\Phi_{a,b}(x_k)=(-1)^{b-1}\frac{r_3(x_k)h(x_k,\bmub)}{f(x_k,\blab)h(\bmuc,x_k)}
+\frac{h(\bmub,x_k)}{h(\bmuc,x_k)}.
\ee
Thus, we see that the limit
$w\to\infty$ of the entries $\tNot_{jk}$ with $j,k\ne a+b+2$ coincides with the entries of the matrix $\Ntt_{jk}$. We arrive
at the following intermediate result:
\be{FF12-interm}
-\lim_{w\to\infty}\tfrac{w}c\;
\mathcal{F}^{(1,2)}_{a,b+1}(z|\blac,\bmuc;\blab,\bmub)=f(z,\blab)f(\bmuc,z)\;\Htt_{a,b}\det_{a+b+2}
(\mathcal{M}_{jk}),
\ee
where
\be{Mjk}
\begin{aligned}
&\mathcal{M}_{jk}=\Ntt_{jk},\qquad j,k=1,\dots,a+b+1,\\
&\mathcal{M}_{j,a+b+2}=\Ntt_{jk}\Bigr|_{x'_k=z},\qquad j=1,\dots,a+b+1,\\
&\mathcal{M}_{a+b+2,k}=\Phi_{a,b}(x_k),\qquad k=1,\dots,a+b+2.
\end{aligned}
\ee

In order to get rid of the last $(a+b+2)$-th row we add to it a linear combination of other rows.
Let
\be{def-Omega}
\begin{array}{l}
{\dis \Omega_j=\frac{g(\lac_j,\blac_j)}{g(\lac_j,\blab)}
,\qquad j=1,\dots,a+1,}\num
{\dis \Omega_{a+1+j}=-\frac{g(\mub_j,\bmub_j)}{g(\mub_j,\bmuc)},}\qquad j=1,\dots,b.
\end{array}
\ee
Then (see appendix~\ref{S-SumF})
\be{sum-N1}
\Phi_{a,b}(x_k)+\sum_{j=1}^{a+b+1}\Omega_j\Ntt_{jk}=\frac{\tau(x_k|\blab,\bmub)-\tau(x_k|\blac,\bmuc)}{f(x_k,\blab)f(\bmuc,x_k)}.
\ee
If $x_k\in\blab$ or
$x_k\in\bmuc$, then due to  Bethe equations the eigenvalues $\tau(x_k|\blab,\bmub)$ and $\tau(x_k|\blac,\bmuc)$ are not singular.
In this case  the corresponding matrix element vanishes due to the factor $f^{-1}(x_k,\blab)f^{-1}(\bmuc,x_k)$. The only non-vanishing element
in the modified last row is the one where $x_k=z$. Therefore, the determinant reduces to the product
of this matrix element and its cofactor, and we arrive at \eqref{FF-ans-13-mod}.

\section{Form factors with infinite Bethe roots\label{S-IBR}}

In this section we consider a special case when one of  the Bethe roots is infinite. As we have seen already, in this case
one should consider renormalized Bethe vectors, for instance,
\be{examples}
\lim_{w\to\infty}w\;\mathbb{B}^{a,b}(\{\bla,w\};\bmu), \qquad
\lim_{w\to\infty}w\;\mathbb{C}^{a,b}(\bla;\{\bmu,w\}),\qquad\text{etc.}
\ee

\subsection{Action of zero modes}
On-shell Bethe vectors with infinite parameters are not necessarily  singular
 weight vectors for the zero modes $T_{ij}[0]$ with
$i>j$. Consider, for example, the  action of $T_{21}[0]$ on the vector $w\;\mathbb{B}^{a,b}(\{\bla,w\};\bmu)$ at $w\to\infty$ and
$\bla$, $\bmu$ finite.
Due to \eqref{act-12} we have
\be{act-21-inf}
T_{21}[0]\lim_{w\to\infty}w\;\mathbb{B}^{a,b}(\{\bla,w\};\bmu)= c\;T_{21}[0] T_{12}[0]\mathbb{B}^{a-1,b}(\bla;\bmu),
\ee
where $\mathbb{B}^{a-1,b}(\bla;\bmu)$ is an on-shell vector depending on finite parameters. Setting $i=l=2$, $j=k=1$  in
\eqref{RTT-zero} and taking the limit $u\to\infty$ we obtain
\be{21-12-zero}
\bigl[T_{21}[0], T_{12}[0]\bigr]  =  T_{11}[0]-T_{22}[0].
\ee
Since $T_{21}[0] \mathbb{B}^{a-1,b}(\bla;\bmu)=0$, we finally arrive at
\be{act-21-inf-res}
T_{21}[0]\lim_{w\to\infty}w\;\mathbb{B}^{a,b}(\{\bla,w\};\bmu)= c(T_{11}[0]-T_{22}[0])\mathbb{B}^{a-1,b}(\bla;\bmu)=
c(r_1[0]+b-2a)\mathbb{B}^{a-1,b}(\bla;\bmu),
\ee
where we have used \eqref{act-11} and \eqref{act-22}.

On the other hand, if we consider an on-shell vector $w\;\mathbb{B}^{a,b}(\bla;\{\bmu,w\})$ at $w\to\infty$, we can easily show that it is
annihilated by the zero mode $T_{21}[0]$. Indeed,
\be{act-21-inf1}
T_{21}[0]\lim_{w\to\infty}w\;\mathbb{B}^{a,b}(\bla;\{\bmu,w\})= c\;T_{21}[0] T_{23}[0]\mathbb{B}^{a,b-1}(\bla;\bmu).
\ee
It follows from \eqref{RTT-zero} that $\bigl[T_{21}[0], T_{23}[0]\bigr]=0$. Then, the zero mode $T_{21}[0]$ acts on the on-shell
Bethe vector $\mathbb{B}^{a,b-1}(\bla;\bmu)$ depending on finite parameters, and we arrive at
\be{act-21-inf2}
T_{21}[0]\lim_{w\to\infty}w\;\mathbb{B}^{a,b}(\bla;\{\bmu,w\})= 0.
\ee
Thus, the result of the action of the zero modes on on-shell Bethe vectors with infinite parameter depends on which set ($\bla$ or $\bmu$) contains  this
infinite argument. We have seen that the action of $T_{21}[0]$ gives non-vanishing result, if the infinite argument belongs to the set $\bla$. Similarly,
one can show that the zero mode $T_{32}[0]$ does not annihilate on-shell vector, if the infinite argument belongs to the set $\bmu$.

It is clear that on-shell dual Bethe vectors with infinite parameters possess analogous properties, namely, they are not
always singular
weight vectors for the zero modes $T_{ij}[0]$ with $i<j$. This statement follows from the results described above and
the mapping \eqref{psiBV}.   In particular, in the next section we will use
\be{act-23-inf}
\lim_{w\to\infty}w\;\mathbb{C}^{a,b}(\bla;\{\bmu,w\}) T_{23}[0]=
c(a-2b-r_3[0])\mathbb{C}^{a,b-1}(\bla;\bmu).
\ee

\subsection{Form factor of $T_{13}(z)$ with infinite Bethe root\label{SS-FFIBR}}

Determinant representation \eqref{FF-ans-13-mod} for the form factor $\mathcal{F}_{a,b}^{(1,3)}(z)$ was obtained under  the assumption
of finiteness of the Bethe roots. Consider now the form factor $\mathcal{F}_{a,b}^{(1,3)}(z)$ depending on an infinite Bethe parameter.
Let, for instance, $\muc_{b+1}\to\infty$. Then
\be{ff-def}
\lim_{\muc_{b+1}\to\infty} \muc_{b+1}\;\mathcal{F}_{a,b}^{(1,3)}(z|\blac, \bmuc;\blab,\bmub)=
\lim_{\muc_{b+1}\to\infty} \muc_{b+1}\;\mathbb{C}^{a+1,b+1}(\blac;\bmuc) T_{13}(z) \mathbb{B}^{a,b}(\blab;\bmub).
\ee
There exist at least two ways to compute this limit. First, due to \eqref{act-32d} we  rewrite \eqref{ff-def} as follows:
\be{ff-rew}
\lim_{\muc_{b+1}\to\infty} \muc_{b+1}\;\mathcal{F}_{a,b}^{(1,3)}(z|\blac, \bmuc;\blab,\bmub)=
c\;\mathbb{C}^{a+1,b}(\blac; \bmuc_{b+1})T_{32}[0] T_{13}(z) \mathbb{B}^{a,b}(\blab;\bmub).
\ee
From \eqref{RTT-zero} we find
\be{com-3213}
[T_{32}[0],T_{13}(z)]= T_{12}(z),
\ee
and using $T_{32}[0] \mathbb{B}^{a,b}(\blab;\bmub)=0$, we finally arrive at
\begin{equation}\label{ff-fin}
\lim_{\muc_{b+1}\to\infty} \muc_{b+1}\;\mathcal{F}_{a,b}^{(1,3)}(z|\blac, \bmuc;\blab,\bmub)
=c\;\mathcal{F}_{a,b}^{(1,2)}(z|\blac, \bmuc_{b+1};\blab,\bmub).
\end{equation}
Thus, we see that the form factor $\mathcal{F}_{a,b}^{(1,3)}(z)$  reduces to $\mathcal{F}_{a,b}^{(1,2)}(z)$ at $\muc_{b+1}\to\infty$.
Using explicit determinant formulas for these two form factors given in section~\ref{S-FFT13} one can check \eqref{ff-fin} directly
(see appendix~\ref{A-DC}).

Another way to compute the limit \eqref{ff-def} is to use the formula \eqref{T13-z1}. If $\muc_{b+1}\to\infty$, then this equation takes the form
\begin{multline}
\lim_{\muc_{b+1}\to\infty} \muc_{b+1}\;\mathbb{C}^{a+1,b+1}(\blac;\bmuc) T_{13}(z)\mathbb{B}^{a,b}(\blab;\bmub)\\
= -\lim_{\muc_{b+1}\to\infty} \muc_{b+1}\;\Bigl(\mathbb{C}^{a+1,b+1}(\blac;\bmuc) T_{12}(z)T_{23}[0]\mathbb{B}^{a,b}(\blab;\bmub) \\
-\mathbb{C}^{a+1,b+1}(\blac;\bmuc) T_{23}[0]T_{12}(z)\mathbb{B}^{a,b}(\blab;\bmub)\Bigr).
\label{T13-zinf}
\end{multline}
In distinction of \eqref{T13-z1}, now the action of the zero mode $T_{23}[0]$ to the left gives non-vanishing contribution due to
\eqref{act-23-inf}. We obtain
\begin{multline}\label{FF-FF-inf}
\lim_{\muc_{b+1}\to\infty} \muc_{b+1}\;\mathcal{F}^{(1,3)}_{a,b}(z|\blac,\bmuc;\blab,\bmub)=
-\lim_{\substack{\muc_{b+1}\to\infty\\ w\to\infty}}\tfrac{w\muc_{b+1}}c\;
\mathcal{F}^{(1,2)}_{a,b+1}(z|\blac,\bmuc;\blab,\{\bmub,w\})\\
+c(a-2b-r_3[0])\mathcal{F}^{(1,2)}_{a,b}(z|\blac,\bmuc_{b+1};\blab,\bmub).
\end{multline}
It seems that we come to a contradiction with \eqref{FF-FF}. Indeed, multiplying \eqref{FF-FF} by $\muc_{b+1}$ and taking
the limit $\muc_{b+1}\to\infty$ we obtain only the first line of \eqref{FF-FF-inf}, without the additional term
in the second line of this equation.

The reason of this apparent contradiction is due to a subtlety  hidden in the structure of the  determinant representations for the form factors.
We shall describe this subtlety in details for representation \eqref{FF-ans-m} of the form factor $\mathcal{F}^{(1,2)}_{a,b+1}(z)$. We would like to
mention, however, that the determinant formulas for all other form factors possess the same properties.

The entries of the matrix $\Not_{jk}$  are given explicitly in \eqref{Lu-m}--\eqref{Lw-m}. Observe that they depend on the functions
$r_1(\lab_k)$ and $r_3(\muc_k)$. Since the sets $\blab$ and $\bmuc$ satisfy  the Bethe equations, one can replace these functions by  products
of the functions $f$ via \eqref{AEigenS-1}. However it would be a mistake to make this replacement without an additional specification
on Bethe parameters.
Formally, equations \eqref{Lu-m}--\eqref{Lw-m} are valid when the Bethe parameters of the vectors $\mathbb{C}^{a+1,b+1}(\blac;\bmuc)$ and $\mathbb{B}^{a,b+1}(\blab;\{\bmub,w\})$ are different, i.e. $\blac\cap\blab=\emptyset$ and $\bmuc\cap\{\bmub,w\}=\emptyset$. If some of them coincide, i.e.
$\blac\cap\blab\ne\emptyset$ or $\bmuc\cap\{\bmub,w\}\ne\emptyset$, then formulas \eqref{Lu-m}--\eqref{Lw-m} remain correct, but one should take
the corresponding limits (see e.g. \cite{BelPRS12b}). In this case one first should take the limit and only after this, one can express the functions
$r_1(\lab_k)$ and $r_3(\muc_k)$  through the Bethe equations. The reverse procedure is incorrect, because we cannot consider a limit
where one solution of Bethe equations goes to another.

Let $\muc_{b+1}=w$ in \eqref{Lw-m}. Then the matrix element $\Not_{a+b+2,a+b+2}$ has a pole. It is easy to see that due to
the Bethe equations \eqref{AEigenS-1} the residue in this pole vanishes, hence, the limit of $\Not_{a+b+2,a+b+2}$ is finite. Taking
the limit $\muc_{b+1}\to w$, we obtain
\be{diag-el}
\Not_{a+b+2,a+b+2}=c\frac{h(\bmub,w)}{h(\bmuc,w)}\left[\frac{r'_3(w)}{r_3(w)}-
\sum_{i=1}^b\frac{2c}{(w-\mub_i)^2-c^2}+\frac1c\sum_{j=1}^at(w,\lab_j)\right],
\ee
where $r'_3(w)$ is the derivative of $r_3(w)$. It is this expression for $\Not_{a+b+2,a+b+2}$ that needs to be used in the case
$\muc_{b+1}=w$.

Let us turn back to the analysis of the apparent contradiction between \eqref{FF-FF} and \eqref{FF-FF-inf}. Deriving \eqref{FF-FF} we assumed that $\muc_{b+1}$ was finite. Then we can multiply \eqref{FF-FF} by $\muc_{b+1}$ and take the limit $\muc_{b+1}\to\infty$. In this case we have in the r.h.s. of
\eqref{FF-FF} the {\it successive} limit: first $w\to\infty$, then $\muc_{b+1}\to\infty$. Thus, taking this successive limit  we do not
set $\muc_{b+1}=w$ in the matrix element $\Not_{a+b+2,a+b+2}$. Actually, it means that we simply take the limit $\muc_{b+1}\to\infty$
in the determinant representation for the form factor $\mathcal{F}^{(1,3)}_{a,b}(z|\blac,\bmuc;\blab,\bmub)$. It is shown in appendix~\ref{A-DC} that
this way agrees with \eqref{ff-fin}.

On the other hand, if we use \eqref{FF-FF-inf}, then we  deal with another case. Indeed, in the second line of \eqref{T13-zinf} we actually have the form factor of the operator $T_{12}(z)$ between two vectors depending on infinite parameters:
\begin{multline}
\lim_{\muc_{b+1}\to\infty} \muc_{b+1}\;\mathbb{C}^{a+1,b+1}(\blac;\bmuc) T_{12}(z)T_{23}[0]\mathbb{B}^{a,b}(\blab;\bmub) \\
=\tfrac 1c\Bigl(\lim_{\muc_{b+1}\to\infty} \muc_{b+1}\;\mathbb{C}^{a+1,b+1}(\blac;\bmuc)\Bigr)\, T_{12}(z)\,\Bigl(\lim_{w\to\infty}
w\;\mathbb{B}^{a,b}(\blab;\{\bmub,w\})\Bigr).
\label{2-line}
\end{multline}
Thus, in this case we should identify $\muc_{b+1}$ and $w$. Therefore the double limit in \eqref{FF-FF-inf} should be understood as follows
\begin{equation}\label{FF-FF-inf1}
\lim_{\substack{\muc_{b+1}\to\infty\\ w\to\infty}}\tfrac{w\muc_{b+1}}c\;
\mathcal{F}^{(1,2)}_{a,b+1}(z|\blac,\bmuc;\blab,\{\bmub,w\})
=\lim_{\substack{\muc_{b+1}\to w\\ w\to\infty}}\tfrac{w^2}c\;
\mathcal{F}^{(1,2)}_{a,b+1}(z|\blac,\bmuc;\blab,\{\bmub,w\}).
\end{equation}
Hence, in this case we have to use the expression \eqref{diag-el} for the matrix element $\Not_{a+b+2,a+b+2}$. It is easy to see
that
\be{diag-el-lim}
\lim_{w\to\infty}\tfrac{w^2}c\Not_{a+b+2,a+b+2}=c(a-2b-r_3[0]).
\ee
Pay attention that this limit exactly coincides with the prefactor in the second line of \eqref{FF-FF-inf}. It is easy to check that
the contribution of \eqref{diag-el-lim} to the determinant of the matrix $\Not$ eventually cancels the additional term in  \eqref{FF-FF-inf}. Thus, \eqref{FF-FF-inf} gives the same result as \eqref{ff-fin} and the apparent contradiction is resolved.

Summarizing all above we conclude that in spite of the determinant representation \eqref{FF-ans-13-mod} for the form factor $\mathcal{F}_{a,b}^{(1,3)}(z)$ was obtained for finite Bethe parameters, it remains valid for  infinite Bethe parameters as well.

\section{Generalization to $GL(N)$ models\label{sect:glN}}
The generalization to models with $GL(N)$ symmetry is rather straightforward.
The $R$-matrix keeps the form \eqref{R-mat} but acts now in auxiliary spaces $V\sim\mathbb{C}^N$.
It commutes with a full $GL(N)$ algebra, generated by the zero modes $T_{j,k}[0]$, $j,k=1,...,N$.
The proofs being identical to the ones given in the $GL(3)$ case, we do not repeat them and just enumerate the properties.
\subsection{Bethe vectors}
Bethe vectors of $GL(N)$ models depend on $N-1$ sets of  parameters $\bar t^{(j)} = \{ t^{(j)}_1,t^{(j)}_2,...,t^{(j)}_{a_j}\}$, $j=1,2,...,N-1$
and $N-1$ integers $a_j$ that correspond to the cardinalities of each set. The Bethe vectors will be noted
\be{}
\mathbb{B}^{\bar a}(\bar t) = \mathbb{B}^{a_1,a_2,...,a_{N-1}}\big(\bar t^{(1)},\bar t^{(2)},...,\bar t^{(N-1)}\big).
\ee
It has been proved in \cite{TarVar93} that they are  singular weight vectors of the $GL(N)$ algebra:
\begin{equation}
\label{act-12-23-gln}
T_{j+1,j}[0]\,\mathbb{B}^{\bar a}(\bar t)=0\,,\quad\mathbb{C}^{\bar a}(\bar t)\,T_{j,j+1}[0]=0,\quad j=1,2,...,N-1,
\end{equation}
if $\mathbb{B}^{\bar a}(\bar t)$and $\mathbb{C}^{\bar a}(\bar t)$ are on-shell Bethe vectors depending on finite parameters.

Then, everything follows the same step as for $GL(3)$. In particular, one can show:
\begin{align}
\label{act-12-gln}
T_{j,j+1}[0]\,\mathbb{B}^{\bar a}(\bar t)&=
\lim_{w\to\infty} \tfrac wc\;
\mathbb{B}^{a_1,a_2,...,a_j+1,...,a_{N-1}}\big(\bar t^{(1)},\bar t^{(2)},...,\bar t^{(j-1)},\{w,\bar t^{(j)}\},\bar t^{(j+1)},...,\bar t^{(N-1)}\big).
\end{align}

\subsection{Form factors}
We define the form factors as
\be{def:FF}
\mathcal{F}^{(j,k)}_{\bar a,\bar b}(z|\bar s; \bar t) = \mathbb{C}^{\bar a}(\bar s)\,T_{j,k}(z)\,\mathbb{B}^{\bar b}(\bar t),
\ee
where $\mathbb{C}^{\bar a}(\bar s)$ and $\mathbb{B}^{\bar b}(\bar t)$ are on-shell Bethe vectors, satisfying the Bethe equations:
\be{BE-gln}
\frac{\lambda_i(t^{(i)}_j)}{\lambda_{i+1}(t^{(i)}_j)}=(-1)^{a_i-1}
\prod_{\substack{m=1\\ m\neq j}}^{a_i}
\frac{f(t^{(i)}_j,t^{(i)}_m)}{f(t^{(i)}_m,t^{(i)}_j)}\
\prod_{m=1}^{a_{i-1}}
f(t^{(i)}_j,t^{(i-1)}_m)^{-1}
\prod_{m=1}^{a_{i+1}}
f(t^{(i+1)}_m,t^{(i)}_j)\,,\ \begin{array}{l} 1\leq j\leq a_i\\[1ex] 1\leq i\leq N-1. \end{array}
\ee
We obtain
\be{FF-FF-gln}
\mathcal{F}^{(j,j+1)}_{\bar a,\bar b}(z|\bar s; \bar t)=(-1)^\epsilon\lim_{s^{(j)}_{a_j+1}\to\infty}\tfrac{s^{(j)}_{a_j+1}}c\;
\mathcal{F}^{(j+\epsilon,j+\epsilon)}_{\bar a',\bar b}(z|\bar s'; \bar t)\,,\quad \epsilon=0,1,
\ee
and
\be{F11F22-F12-gln}
\mathcal{F}^{(j,j)}_{\bar a',\bar b}(z|\bar s; \bar t)-
\mathcal{F}^{(j+1,j+1)}_{\bar a',\bar b}(z|\bar s; \bar t)
=\lim_{s^{(j)}_{a_j+1}\to\infty}\tfrac{s^{(j)}_{a_j}}c\;
\mathcal{F}^{(j,j+1)}_{\bar a',\bar b}(z|\bar s'; \bar t).
\ee
We  use the notation $\bar a'=\{a_1,a_2,...,a_j+1,...,a_{N-1}\}$,
$\bar s' = \{ \bar s^{(1)},\bar s^{(2)},...,\bar s^{(j)'},...,\bar s^{(N-1)}\}$ and
$\bar s^{(j)'} = \{ s^{(j)}_1,s^{(j)}_2,...,s^{(j)}_{a_j},s^{(j)}_{a_j+1}\}$.

Finally, from the commutation relation $[T_{j,j+1}[0],T_{j+1,k}(z)]=T_{j,k}(z)$ for $k>j+1$, we get the recursion
\be{FFF-gln}
\mathcal{F}^{(j,k)}_{\bar a,\bar b}(z|\bar s; \bar t)=(-1)^\epsilon\lim_{s^{(j)}_{a_j+1}\to\infty}\tfrac{s^{(j)}_{a_j+1}}c\;
\mathcal{F}^{(j+1,k)}_{\bar a',\bar b}(z|\bar s'; \bar t)\,,\quad 1\leq j<k<N.
\ee
Using the antimorphism $\psi$, we can get similar result for $\mathcal{F}^{(j,k)}_{\bar a,\bar b}(z|\bar s; \bar t)$ with $j>k$.
Thus, we arrive at the following
\begin{prop}\label{all-one-gln}
All form factors in the $GL(N)$-invariant generalized model can be obtained from only one form factor by sending one of Bethe parameters to infinity.
\end{prop}
The problem, however, is to find an appropriate representation for this initial form factor. It remains an open question.

Nevertheless, one can still get some properties for the form factors. For instance, similarly to the $GL(3)$ case we have:
%
\begin{prop}
Given sets $\bar s$ and $\bar t$, let us assume that $\exists\, \ell$ such that $\bar s^{(\ell)}\ne\bar t^{(\ell)}$. Then for all form factors
$\mathcal{F}^{(j,k)}_{\bar a,\bar b}(z|\bar s;\bar t)$ the dependence on the external parameter $z$ is given by
\be{dep-z1}
\mathcal{F}^{(j,k)}_{\bar a,\bar b}(z|\bar s;\bar t)=\bigl(\tau(z|\bar s)-\tau(z|\bar t)\bigr)\cdot
\UF^{(j,k)}_{\bar a,\bar b}(\bar s;\bar t),
\ee
where $\tau(z|\bar t)$ is the transfer matrix eigenvalue
\be{tau-gln}
\tau(z;\bar{t})
=\sum_{i=1}^N \lambda_i(z)\,\prod_{j=1}^{a_{i-1}}f(z,t^{(i-1)}_j)\,
\prod_{j=1}^{a_{i}}f(t^{(i)}_j,z)\,.
\ee
The  universal form factor
$\UF^{(j,k)}_{\bar a,\bar b}(\bar s;\bar t)$ does not depend on $z$.
\end{prop}

\section*{Conclusion}

In this paper we have developed a new method of calculation of form factors of the monodromy matrix entries
in $GL(3)$-invariant
integrable models. The method is based on the use of the zero modes in the expansion  of the monodromy matrix.
We obtained determinant representations
for all form factors $\mathcal{F}_{a,b}^{(i,j)}(z)$ and showed that they are related to each other. In particular, we have proved that
all the form factors can be obtained from the initial one by taking special limits of the Bethe parameters.

The obtained results can be used for the calculation of form factors and correlation functions in the $SU(3)$-invariant $XXX$
Heisenberg chain. For this model the solution of the quantum inverse scattering problem is known \cite{MaiT00}. Therefore form
factors of local operators in the $SU(3)$-invariant $XXX$ Heisenberg chain can be easily reduced to the ones considered in
the present paper.

However this is not the only possible application of the determinant formulas for form factors and the method of the zero modes.
The last one opens a new way to study form factors and correlation functions in other quantum models solvable by the nested algebraic
Bethe ansatz. In particular, this method can be applied to the model of two-component one-dimensional gases with $\delta$-function
interaction \cite{Yang67,Sath68,PozOK12}. We are planning to attack this problem in our forthcoming publications.

The calculation of form factors for models with $GL(N)$ symmetry remains to be done.
Obviously, a determinant form is far from being achieved, but the zero modes method reduces the `quest' to only
one form factor, or even to the scalar product of Bethe vectors.

Another natural question deals with models of $XXZ$ type. At that point, it is not clear to us whether
the zero modes method can be applied in this context. In particular, on-shell Bethe vectors (with non-infinite Bethe parameters)
are no longer singular weight vectors. Since it is an essential property to deduce some of the
relations we used, it may be an indication that one cannot extend directly the zero modes method to the $XXZ$ type models.
We will address this problem in our future publications.

\section*{Acknowledgements}
The work of S.P. was supported in part by RFBR grant 14-01-00474-a. E.R. was supported by ANR Project
DIADEMS (Programme Blanc ANR SIMI1 2010-BLAN-0120-02).
N.A.S. was  supported by the Program of RAS  ``Nonlinear Dynamics in Mathematics and Physics'',
RFBR-14-01-00860-a, RFBR-13-01-12405-ofi-m2.

\appendix

\section{Summation formulas\label{S-SumF}}

\begin{prop}\label{Omega-sum}
 Let $\#\blac=a+1$, $\#\blab=a$, $\#\bmuc=b+1$ , and $\#\bmub=b$. Let  an $(a+b+1)$-component vector $\Omega$ be
as  in \eqref{def-Omega}. Then
\be{tuz}
\begin{aligned}
\sum_{j=1}^{a+1}t(\lac_j,z)\Omega_j&=\frac{h(\blab,z)}{h(\blac,z)}\left(\frac{f(\blac,z)}{f(\blab,z)}-1\right),\\
\sum_{j=1}^{ a+1}t(z,\lac_j)\Omega_j&=\frac{h(z,\blab)}{h(z,\blac)}\left(\frac{f(z,\blac)}{f(z,\blab)}-1\right),\\
\sum_{j=1}^bt(\mub_j,z)\Omega_{a+1+j}&=\frac{h(\bmuc,z)}{h(\bmub,z)}\left(1-\frac{f(\bmub,z)}{f(\bmuc,z)}\right)-1,\\
\sum_{j=1}^bt(z,\mub_j)\Omega_{a+1+j}&=\frac{h(z,\bmuc)}{h(z,\bmub)}\left(1-\frac{f(z,\bmub)}{f(z,\bmuc)}\right)-1.
\end{aligned}
\ee
\end{prop}
{\sl Proof}. All the identities above can be proved in the same way. Consider, for example, the third identity.
Let
 \be{sim-sum}
 \sum_{j=1}^b t(\mub_j,z)\Omega_{a+1+j}=W(z).
\ee
The sum in the l.h.s. of \eqref{sim-sum} can be computed by means of an auxiliary integral
\be{Integ}
I=\frac{-1}{2\pi i}\oint\limits_{|\omega|=R\to\infty}
\frac{d\omega}{(\omega-z)(\omega-z+ c)}
\frac{\prod_{\ell=1}^{b+1} (\omega-\muc_\ell)}{\prod_{\ell=1}^{b}(\omega-\mub_\ell)}.
\ee
The integral is taken over the anticlockwise oriented contour $|\omega|=R$ and we consider the limit $R\to\infty$. Then $I=-1$, because the integrand behaves as $1/\omega$ at $\omega\to\infty$. On the other hand the same integral is equal to the sum
of the residues within the integration contour. Obviously the sum of the residues at $\omega=\mub_\ell$ gives
$W(z)$. There are also two additional poles at $\omega=z$ and $\omega=z-c$. Then we have
 \be{I-res}
 I=-1=W(z)+ \frac1c\,\frac{\prod_{\ell=1}^{b+1} (z-\muc_\ell-c)}{\prod_{\ell=1}^{b}(z-\mub_\ell-c)}
 -\frac1c\,\frac{\prod_{\ell=1}^{b+1} (z-\muc_\ell)}{\prod_{\ell=1}^{b}(z-\mub_\ell)}.
 \ee
From this we obtain the third identity \eqref{tuz}.

Using identities \eqref{tuz} we can easily derive
\be{sum-N}
\sum_{j=1}^{a+b+1}\Omega_j\Ntt_{jk}=\frac{\tau(x_k|\blac,\bmuc)-\tau(x_k|\blab,\bmub)}{f(x_k,\blab)f(\bmuc,x_k)}-\Phi_{a,b}(x_k).
\ee

\section{Direct check of \eqref{ff-fin}\label{A-DC}}

In order to check \eqref{ff-fin} it is convenient to use representation \eqref{FF12-interm}. Let us introduce
\be{tH13}
\tHtt_{a,b}=\left(\tfrac{\muc_{b+1}}{c}\right)^{-b}\Htt_{a,b},
\ee
and a matrix $\tNtt$
\be{tM}
\begin{aligned}
\tNtt_{jk}&=\mathcal{M}_{jk},\qquad &j=1,\dots,a+1,\qquad k=1,\dots,a+b+2,\\
\tNtt_{a+1+j,k}&=\tfrac{\muc_{b+1}}c\;\mathcal{M}_{a+1+j,k},\qquad &j=1,\dots,b+1,\qquad k=1,\dots,a+b+2.
\end{aligned}
\ee
Here $\Htt_{a,b}$ is given by \eqref{hHab}, the matrix $\mathcal{M}$ is given by \eqref{Mjk}. Then taking into account \eqref{FF12-interm}
we recast equation \eqref{ff-fin} in the form
\begin{equation}\label{ff-fin1}
\mathcal{F}_{a,b}^{(1,2)}(z|\blac, \bmuc_{b+1};\blab,\bmub)=\lim_{\muc_{b+1}\to\infty} f(z,\blab)f(\bmuc,z)\;\tHtt_{a,b}\det_{a+b+2}
(\tNtt_{jk}).
\end{equation}

It is easy to see that
\be{lim-tHtt}
\lim_{\muc_{b+1}\to\infty} f(z,\blab)f(\bmuc,z)\;\tHtt_{a,b}=\Hot_{a,b}.
\ee
Consider the limit $\muc_{b+1}\to\infty$ of the entries $\tNtt_{jk}$. In the $(a+b+1)$-th column of this matrix
$x_k=\muc_{b+1}$ and one can easily find that
\be{lim-last}
\lim_{\muc_{b+1}\to\infty}\tNtt_{j,a+b+1}=0,\qquad j=1,\dots,a+b+1.
\ee
Thus, the determinant of $\tNtt$ reduces to the product of $\tNtt_{a+b+2,a+b+1}$ by the corresponding cofactor, where we have
\be{lim-cofac}
\lim_{\muc_{b+1}\to\infty}\tNtt_{jk}=\Not_{jk},\qquad j\ne a+b+2, \qquad k\ne a+b+1.
\ee
Thus, we arrive at
\begin{multline}\label{lim-interm}
\lim_{\muc_{b+1}\to\infty} f(z,\blab)f(\bmuc,z)\;\tHtt_{a,b}\det_{a+b+2}
(\tNtt_{jk})\\
=-\Hot_{a,b}\det_{a+b+1}(\Not_{jk})\cdot\lim_{\muc_{b+1}\to\infty}\tNtt_{a+b+2,a+b+2}\;.
\end{multline}
The element $\tNtt_{a+b+2,a+b+2}$ is equal to
\be{M-phi}
\tNtt_{a+b+2,a+b+2}= \tfrac{\muc_{b+1}}c\;\Phi_{a,b}(\muc_{b+1})=\tfrac{\muc_{b+1}}c\;\Bigl((-1)^{b-1}\frac{r_3(\muc_{b+1})h(\muc_{b+1},\bmub)}{f(\muc_{b+1},\blab)h(\bmuc,\muc_{b+1})}
+\frac{h(\bmub,\muc_{b+1})}{h(\bmuc,\muc_{b+1})}\Bigr).
\ee
Expressing $r_3(\muc_{b+1})$ via Bethe equations we obtain
\be{M-phi1}
\tNtt_{a+b+2,a+b+2}=\tfrac{\muc_{b+1}}c\;\Bigl(
\frac{h(\bmub,\muc_{b+1})}{h(\bmuc,\muc_{b+1})}-\frac{f(\muc_{b+1},\blac)h(\muc_{b+1},\bmub)}{f(\muc_{b+1},\blab)h(\muc_{b+1},\bmuc)}\Bigr).
\ee
In order to take the limit $\muc_{b+1}\to\infty$ it is useful to write all the products in \eqref{M-phi1} explicitly
\begin{multline}\label{M-phi2}
\tNtt_{a+b+2,a+b+2}=\tfrac{\muc_{b+1}}c\;\Biggl(
\prod_{i=1}^b\frac{\mub_i-\muc_{b+1}+c}{\muc_i-\muc_{b+1}+c}\\
-\prod_{i=1}^b
\frac{\muc_{b+1}-\mub_i+c}{\muc_{b+1}-\muc_i+c}\prod_{j=1}^{a+1}\frac{\muc_{b+1}-\lac_j+c}{\muc_{b+1}-\lac_j}
\prod_{j=1}^{a}\frac{\muc_{b+1}-\lab_j}{\muc_{b+1}-\lab_j+c}\Biggr).
\end{multline}
Then we find
\be{Lim-Phi}
\lim_{\muc_{b+1}\to\infty}\tNtt_{a+b+2,a+b+2}=-1,
\ee
and we finally arrive at
\begin{equation}\label{lim-fin}
\lim_{\muc_{b+1}\to\infty}  f(z,\blab)f(\bmuc,z)\;\tHtt_{a,b}\det_{a+b+2}
(\tNtt_{jk})=\Hot_{a,b}\det_{a+b+1}(\Not_{jk}),
\end{equation}
in complete agreement with \eqref{ff-fin}.

\end{document}